\documentclass[%
 reprint,%
 amssymb, amsmath,%
 aip,
]{revtex4-1}

\usepackage{float}
\usepackage{graphicx}
\usepackage{bm}%
\usepackage[colorlinks]{hyperref}
\usepackage{wrapfig}
\renewcommand{\vec}[1]{{\bf #1}}
\newcommand{\braket}[1]{\langle #1  \rangle} 
\newcommand{\ket}[1]{| #1  \rangle} 
\newcommand{\bra}[1]{\langle #1 |} 
\newcommand{\Tr}[1]{\text{Tr } #1} 
\newcommand{\abs}[1]{| #1|} 

\draft 

\begin{document}

\title{Quantum Momentum Distribution and Quantum Entanglement in the Deep Tunneling Regime}
\author{Yantao Wu}%
\email{yantaow@princeton.edu}
\affiliation{%
The Department of Physics, Princeton University
}%

\author{Roberto Car}%
\email{rcar@princeton.edu}
\affiliation{%
The Department of Chemistry and the Department of Physics, Princeton University
}%
\date{\today}
\begin{abstract}
In this paper, we consider the momentum operator of a quantum particle directed along the displacement of two of its neighbors. 
A modified open-path path integral molecular dynamics is presented to sample the distribution of this directional momentum distribution, where we derive and use a new estimator for this distribution.  
Variationally enhanced sampling is used to obtain this distribution for an example molecule, Malonaldehyde, in the very low temperature regime where deep tunneling happens.
We find no secondary feature in the directional momentum distribution, and that its absence is due to quantum entanglement through a further study of the reduced density matrix.  
\end{abstract}

\maketitle

\section{Introduction}
Quantum mechanical phenomena, such as zero-point motion and tunneling, affect the
equilibrium configurations of molecules and materials containing light atoms up to room
temperature and above. These effects are ignored in classical atomistic simulations but
are accessible to methodologies like path integral molecular dynamics (PIMD), which
are based on Feynman path sampling \cite{Feynman_PI}. 
In the most common applications of this
technique, one is interested in position dependent observables when the
indistinguishability of identical quantum particles can be ignored. Then, the relevant
equilibrium averages can be evaluated via closed-path PIMD sampling. 
In this
approach, the Feynman paths in imaginary time that describe the quantum particles are
discretized \cite{PIMD_1, PIMD_2}, and the equilibrium statistical averages are calculated by sampling with
molecular dynamics an appropriate classical system of ring polymers. 
Each polymer includes $l$ beads labeled by an integer index $i$ varying from 0 to $l$ with the condition that the $i=0$ and the $i=l$ beads coincide.
Often in these
studies zero-point motion was the only relevant quantum effect, but occasionally tunneling situations have also been considered. Modeling tunneling \cite{Tunneling_book} is important as this phenomenon can facilitate chemical reactions and structural phase transitions. 
When the barrier separating two tunneling configurations is of the order of the thermal energy available to a ring polymer, the latter is able to frequently switch back and forth between
the two configurations on the time scale of PIMD sampling, a situation that is often referred to as shallow tunneling regime. 
On the other hand, when the barrier is large we are in the deep tunneling regime in which barrier crossing is infrequent.
In the spite of the sampling difficulties that numerical methods face in presence of a high energy barrier, various studies of systems in the deep tunneling regime have been carried out \cite{tunneling_splitting, TS_water, RPMD_rate, FES_clathrate}.   
It has been suggested that momentum dependent observables such as the particle momentum distribution might exhibit features that are more clearly identifiable with quantum tunneling than space dependent observables \cite{open_path_water}. 
For example, the space distribution of a particle in a double well potential is bimodal, but the bimodality can be
either due to tunneling or to thermal hopping. 
On the other hand, a tunneling particle in
the ground state of a double well potential in one dimension (deep tunneling) would
show a node, i.e. a point of zero value for the distribution, separating two clear features
at zero and finite momentum \cite{kdp, open_path_water}. 
This is very different from the Gaussian momentum
distribution associated to a classical particle. 
If the
tunneling particle was not in the ground state, the mathematical node would 
disappear. 
However, if the quantum state of the particle remained dominated by the two tunnel split states as one expects for the deep tunneling regime, the momentum
distribution would retain a secondary feature.  
This is not an issue of academic interest only, because the momentum distribution of an
atom in a condensed phase environment can be measured with deep inelastic neutron scattering (DINS) experiments \cite{kdp, lightatoms, REITER2004}. Often these measure spherical averages that tend to wash out the tunneling features, but experiments on crystalline samples can measure
the momentum distribution along specific crystallographic directions. 
For example, one
such experiment suggested presence of two features attributed to deep quantum
tunneling in crystalline potassium diphosphate (KDP) \cite{kdp}, a system with a ferroelectric-
paraelectric transition caused by hydrogen atoms undergoing tunneling.
PIMD can provide information on the momentum space via open-path
simulations 
\cite{Open_path_He,Open_path_OH,III,JJJ,KKK}. 

The momentum distribution of one atom, say $A$, is given by $n(\vec p) = \delta(\vec p_A - \vec p) = \Tr(\delta(\hat{\vec p}_A - \vec p) \hat{\rho})$, where the hat indicates quantum mechanical operators,
and $\hat{\rho} = e^{-\beta \hat{H}}/\Tr(e^{-\beta\hat{H}})$ is the full density operator of a system of $N$ atoms with Hamiltonian $\hat{H}$ at
inverse temperature $\beta \equiv (k_BT)^{-1}$. 
The momentum distribution is the Fourier transform, $n(\vec p) = (2\pi\hbar)^{-3} \int d^3\vec x e^{i\vec p\cdot \vec x}n(\vec x)$, of the end-to-end displacement distribution $n(x)$ given by:
\begin{equation}
  \label{eq:nx_normal}
  n(\vec x) = \int d^{3N} \vec r d^{3N} \vec r' 
  \delta(\vec r_{\not=A} - \vec r'_{\not=A}) \delta(\vec r_{A} - \vec r'_{A} - \vec x) \rho(\vec r, \vec r')   
\end{equation}
where $\rho(\vec r, \vec r') = \frac{\braket{\vec r|e^{-\hat{H}}|\vec r'}}{\Tr(e^{-\hat{H}})}$ is the full density matrix of a system of $N$ atoms in the position space representation. 
Here $\vec r_A$ is a three-dimensional position vector of atom $A$ and $\vec r_{\not=A}$ is a $(3N-3)$-dimensional position vector of all the atoms in the system other than $A$.   
Together they make the $3N$-dimensional positive vector of the full system: $\vec r_A \otimes \vec r_{\not=A} = \vec r$. 

$n(\vec x)$ can be evaluated with a PIMD simulation in which the path corresponding to $A$ is kept open, i.e. the two end beads of the corresponding polymer chain are allowed to move freely, while the polymer chains corresponding to all the other atoms are kept closed.  
In an $l$-bead open path PIMD, $\vec x$ can be evaluated with the estimator         
\begin{equation}
  \vec x = \vec r_A^0 - \vec r_A^l    
\end{equation}
where $\vec r_A^i$ is the position of the $i$th bead of $A$. 
In the last decade, open-path PIMD simulations have
been used to compute atomic momentum distributions in molecular and condensed
phase environments \cite{enzyme, open_path_water, linlin}, showing, in particular, that ab-initio PIMD simulations can predict
momentum distributions in good agreement with DINS experiments. 
In these simulations the interatomic interactions are derived from the instantaneous electronic
ground-state within density functional theory. 
One such study investigated the pressure
induced transition between two high-pressure forms of ice, ice VIII and ice VII, in a
temperature regime in which the transition is promoted by quantum tunneling of the
hydrogens participating in the hydrogen bonds \cite{water_jsp}. No bimodal momentum distribution was
found, but the study could only be performed under shallow tunneling conditions due to
the overwhelming computational cost of ab-initio PIMD simulations. 
An interesting result
was that tunneling in high-pressure ice involves a highly correlated motion of several
hydrogens that contributes, by quantum entanglement, to wash out the secondary,
tunneling related, feature of the momentum distribution. 
It would be of interest to
investigate whether this conclusion remains valid in the deep tunneling regime.
Moreover, even in situations in which a single atom participates in tunneling, its motion
in a molecular environment is not strictly one dimensional. Entanglement due to
coupling with the motion of other atoms, could wash out the secondary feature in the
momentum distribution of the tunneling atom, even in absence of correlated motions of
several tunneling particles.
In this paper, to investigate the above issues and make a close comparison between the momentum distribution in one-dimension and the full many-body motion in the deep tunneling regime, we consider in the many-body case the directional momentum distribution, $n(p)$ of atom $A$, projected along the axis defined by the displacement between its two neighboring atoms $B$ and $C$: 
\begin{equation}
  \label{eq:np_dist}
  n(p) = \Tr{(\delta(\hat{\vec p}_A \cdot \frac{\hat{\vec r}_B - \hat{\vec r}_C}{\abs{\hat{\vec r}_B - \hat{\vec r}_C}} - p) \, \cdot \hat{\rho})}
\end{equation}
Our approach combines a general PIMD scheme for sampling the directional momentum distribution, valid for both
molecules and crystals, with variational enhanced sampling (VES) \cite{ves} to overcome the rare
event character of deep tunneling. VES exploits a variational principle to find the optimal
bias potential that facilitates sampling phase space regions separated by energetic
and/or entropic bottlenecks. 
The bias potential depends on suitable collective
coordinates. VES has been used successfully in a number of problems including rare
molecular conformational changes \cite{ves}, nucleation in first order phase transitions \cite{ves_nucleation}, and even scale transformations in real space renormalization group theory \cite{vmcrg}. 
Here we show that it is
also useful to model quantum tunneling when the tunneling time is long on the scale of
molecular dynamics. 
We demonstrate the VES approach to tunneling by applying it to a
relatively simple molecular system, Malonaldehyde \cite{Hbond_Mal} (Fig. \ref{fig:11}), in which a hydrogen atom is known
to tunnel between two equivalent sites \cite{Tunneling_Mal}. 
Tunneling in this molecule has been studied
experimentally \cite{mal_spectroscopy,tunneling_rotation,matrix_quenching,vibrational_tuning} and theoretically \cite{Mal1,Mal2,Mal3,Mal4,Mal5,Mal6,pes,tunneling_splitting, Tunneling_Mal} and there are good estimates for the
tunnel splitting. 
Importantly, the many-body potential energy surface of this molecule
and its analytic gradients, i.e. the forces on the atoms, have been accurately
parametrized and are available \cite{pes}. 
The parametrized potential reproduces well the tunnel
splitting and, indeed, has been used recently in an interesting study of tunnel splitting by
PIMD \cite{tunneling_splitting}. 
The availability of a parametrized potential energy surface means that we
can control accurately the systematic and statistical errors incurred in our PIMD
simulations of the momentum distribution. In addition, we can compare tunneling in the
many-body potential energy surface with that on a one-body potential energy surface in
which the coordinates of all the atoms with the exception of the tunneling hydrogen have
been frozen.
Our main results are the following. 
At a temperature of $63$K, or equivalently inverse temperature $\beta = 5000$ a.u., the momentum
distribution of the frozen one-dimensional system clearly exhibits a secondary shoulder as
expected. 
However, when all the atomic degrees of freedom are left free to move, the
correlations of the molecular motions are sufficient to smooth out the secondary features associated to deep tunneling in the one-dimensional double well
model. 
In addition, to explain the qualitative difference between the one-dimensional and the many body system, we study with VES PIMD an effective reduced density matrix associated to the directional momentum distribution to investigate the mechanisms by which the secondary feature of the momentum distribution in the many body system smears out.     
Although at inverse temperature $\beta = 5000$ a.u., the full density matrix of the Malonaldehyde molecule is dominated by the first two energy states, with dominance of the ground state, the eigenvalues of the reduced density matrix show a significant contribution from higher eigenstates, indicating quantum entanglement, which smears out the secondary feature of $\tilde{n}(x)$, and therefore that of $n(p)$. 
This quantum entanglement might pose a fundamental limitation as to how ``featured'' the directional momentum distribution can be, no matter how low the temperature one is able to attain. 

This paper is organized as follows. 
In Sec. \ref{sec:derivation}, we give the estimator to be sampled in PIMD to compute the directional momentum distribution. 
The modified open-path PIMD and the enhanced sampling technique are described in detail in Sec. \ref{sec:sampling}.    
A numerical example of quantum tunneling of Malonaldehyde is given in Sec. \ref{sec:mal}.
The computation and discussion of the reduced density matrix is given in \ref{sec:rdm}. 
In Sec. \ref{sec:summary}, we summarize our results and discuss possible future work.  
\section{The directional momentum Distribution in Quantum Statistical Mechanics and in PIMD}
\label{sec:derivation}
To compute the directional momentum distribution in Eq. \ref{eq:np_dist}, one needs to sample a well-behaved estimator in a PIMD simulation. 
As proved in \ref{sec:dist_derivation}, the directional momentum distribution, $n(p)$, can be obtained as the Fourier transform of the distribution of a modified end-to-end displacement, $\tilde{n}(x)$:    
\begin{equation}
  \label{eq:np_final}
  \begin{split}
  n(p) &= \Tr (\delta(\hat{\vec p}_A \cdot \frac{\hat{\vec r}_B - \hat{\vec r}_C}{\abs{\hat{\vec r}_B - \hat{\vec r}_C}} - p)\hat{\rho})\\
  &=  \frac{1}{2\pi\hbar} \int\, dx\, e^{ipx}\tilde{n}(x)
\end{split}
\end{equation}
Here $\tilde{n}(x)$ is given by  
\begin{equation}
  \label{eq:nx_final}
  \begin{split}
  \tilde{n}(x) &= \frac{1}{2\pi\hbar}\int\, dx\, e^{ipx} \int d^{3N}\vec r d^{3N}\vec r' \rho(\vec r', \vec r)
  \\
  &\hspace{5mm}\delta(\vec r_{\not=A} - \vec r'_{\not=A}) 
\delta\left(\vec r'_A - \vec r_A + x \cdot \frac{\vec r_B - \vec r_C}{\abs{\vec r_B - \vec r_C}}\right)  
\end{split}
\end{equation}
where the notation $\vec r_{\not=A}, \vec r_A, \vec r_B$, etc., follows that of Eq. \ref{eq:nx_normal}. 
Thus, $\tilde{n}(x)$ can be sampled as a distribution function in a modified form of open-path PIMD, and $n(p)$ can be then obtained.      
\section{Sampling of the directional momentum Distribution with PIMD}
\label{sec:sampling}
Eq. \ref{eq:nx_final} indicates that in order to get the directional momentum distribution of atom $A$ along the displacement of atom $B$ and atom $C$, one should run an open-path PIMD where the polymer chains of all atoms other than $A$ are closed and the polymer chain of atom $A$ is let open along the displacement vector connecting atom $B$ to atom $C$ with an end-to-end distance equal to $x$.  
\subsection{Modified Open-path PIMD}
To derive the equations of motion of the PIMD, we first write $\tilde{n}(x)$ in the form of a path integral:
\begin{equation}
  \label{eq:nx_PI}
    \tilde{n}(x) \propto \underset{\vec r(\beta\hbar) = \vec r(0)}{\int \mathcal{D}[\vec r_{\not=A}(\tau)]} \underset{\vec r_A(\beta\hbar) = \vec r_A(0) - x \cdot \frac{\vec r_B(0) - \vec r_C(0)}{\abs{\vec r_B(0) - \vec r_C(0)}}}{\int \mathcal{D}[\vec r_{A}(\tau)]} e^{-\frac{S[\vec r(\tau)]}{\hbar}}  
\end{equation}
with the action $S[\vec r(\tau)] =\int_0^{\beta\hbar} \frac{1}{2} \sum_{n=1}^N m_n \dot{\vec{r}}_n^2(\tau) + V[\vec r(\tau)]d\tau$.  
To eliminate the awkward factor in the integration boundary of $\mathcal{D}[\vec r_A(\tau)]$, we perform the following change of variable \cite{linlin} $\vec r_A(\tau) \rightarrow  \tilde{\vec r}_A(\tau)$:
\begin{equation}
  \label{eq:r_tilde}
  \vec r_A(\tau) = \tilde{\vec r}_A(\tau) - y(\tau) \cdot x \cdot \frac{\vec r_B(0) - \vec r_C(0)}{\abs{\vec r_B(0) - \vec r_C(0)}}  
\end{equation}
where $y(\tau) = \frac{\tau}{\beta\hbar} - \frac{1}{2}$ so that $\tilde{\vec r}(0) = \tilde{\vec r}(\beta \hbar)$. 
Then one has (see Sec. \ref{sec:transform} for a proof)   
\begin{equation}
  \label{eq:n(x)}
  \tilde{n}(x) \propto \underset{\tilde{\vec r}(\beta\hbar) = \tilde{\vec r}(0)}{\int \mathcal{D}[\tilde{\vec r}(\tau)]} \hspace{1mm} e^{-\frac{S[\tilde{\vec r}(\tau), x]}{\hbar}} \hspace{1mm} e^{-\frac{1}{2} \frac{m_A}{\beta\hbar^2} x^2}  
\end{equation}
with the action 
\begin{equation}
  \begin{split}
    S[ \tilde{\vec r} &(\tau),  x] = \int_0^{\beta\hbar} d\tau \frac{1}{2} \sum_{n=1}^N m_n \dot{\tilde{\vec{r}}}_n^2(\tau) \\ 
  &+ \int_0^{\beta\hbar}d\tau V[\tilde{\vec r}_A(\tau) - x \cdot y(\tau) \cdot \frac{\tilde{\vec r}_B(0) - \tilde{\vec r}_C(0)}{\abs{\tilde{\vec r}_B(0) - \tilde{\vec r}_C(0)}}, \tilde{\vec r}_{\not=A}(\tau)]   
\end{split}
\end{equation}
Discretizing the imaginary time interval $[0, \beta\hbar]$ in terms of $l$ blocks, we can then write down the Hamiltonian of the $l$-bead modified open-path PIMD at inverse temperature $\beta$ \cite{PIMD_1}: 
\begin{equation}
\label{eq:hamiltonian}
\begin{split}
  H( \{\tilde{\vec r}_n^i\}, \{\tilde{\vec p}_n^i\}, x) &= \sum_{n=1}^N\sum_{i=0}^{l-1} \frac{1}{2} m_n \omega_l^2 (\tilde{\vec r}_n^i - \tilde{\vec r}_n^{i+1})^2  \\ 
  & + \frac{1}{l} \sum_{i=1}^{l-1} V[\tilde{\vec r}_A^i - x \cdot y_i \cdot \frac{\tilde{\vec r}_B^0 - \tilde{\vec r}_C^0}{\abs{\tilde{\vec r}_B^0 - \tilde{\vec r}_C^0}}, \tilde{\vec r}_{\not=A}^i]  \\
  & + \frac{1}{2l} \hspace{1mm} V[\tilde{\vec r}_A^0 - x\cdot y_0 \cdot \frac{\tilde{\vec r}_B^0 - \tilde{\vec r}_C^0}{\abs{\tilde{\vec r}_B^0 - \tilde{\vec r}_C^0}}, \tilde{\vec r}_{\not=A}^0]  \\
  & + \frac{1}{2l} \hspace{1mm} V[\tilde{\vec r}_A^0 - x\cdot y_l \cdot \frac{\tilde{\vec r}_B^0 - \tilde{\vec r}_C^0}{\abs{\tilde{\vec r}_B^0 - \tilde{\vec r}_C^0}}, \tilde{\vec r}_{\not=A}^0] \\ 
  & + \frac{1}{2} \frac{m_A}{(\beta\hbar)^2}x^2 + K(\{\tilde{\vec p}_n^i\}, x) 
\end{split}
\end{equation}
where $y_i = \frac{i}{l} - \frac{1}{2}$, for $i = 0, \cdots, l$.  
$\omega_l = \frac{\sqrt{l}}{\beta\hbar}$, $\tilde{\vec r}_n^i$ is the position of the $i$th bead of atom $n$ and $\tilde{\vec r}_n^{l} = \tilde{\vec r}_n^{0}$ in the first term for all $n$. 
The first term on the right-hand side of Eq. \ref{eq:hamiltonian} represents the harmonic potential energy of the beads, in which $m_n$ is the mass of atom $n$, $V$ is the potential energy associated to the many-body
interaction between the atoms, $K(\{\tilde{\vec p}_n^i\}, x)$ is the classical kinetic energy of the beads in which
$\tilde{\vec p}_n^i$ is the MD momentum of the $i$th bead associated to the $n$th atom.
The masses in the kinetic energy $K(\{\tilde{\vec p}_n^i\}, x)$ can be chosen freely.
In this paper, we choose them to be the physical masses of the atoms. 

In the $l$-bead PIMD that we have implemented there are $3Nl + 1$ degrees of freedom: $Nl$ beads in three dimensions and the end-to-end displacement, $x$, which describes the constrained position of the $l$th bead of $A$. (Here we denote the starting bead of a path-integral polymer chain as the 0th bead.) 
To simulate the equation of motion, we note that the quadratic part of $H$ can be integrated analytically in the same way as in a common closed-path PIMD.    
In the velocity Verlet algorithm \cite{v-verlet}, which we use, one should place this quadratic part in the inner loop of the Trotter splitting of the MD integrator, and evolve the MD momentum with the combined action of the potential $V$ and a thermostat in the outer loops. 
Thus, the MD time step $dt$ includes the following updates:
\begin{enumerate}
\item The momentum of the system is propagated by $dt/2$ by the action of the thermostat. 
\item The system momentum is propagated by $dt/2$ by the action of the potential energy $V(\vec r)$: 
  \begin{equation}
    \tilde{\vec p} \rightarrow \tilde{\vec p} - \frac{\partial V(\tilde{\vec r})}{\partial \tilde{\vec r}}dt/2 
  \end{equation}
\item The system momentum and position are propagated analytically by $dt$ with harmonic part of the Hamiltonian. 
\item Step 2 is repeated 
\item Step 1 is repeated 
\end{enumerate}

The distribution of $x$ in the MD run under $H$, when properly thermostatted, is then the $\tilde{n}(x)$ that we seek in Eq. \ref{eq:np_final}.  
\subsection{Enhanced Sampling at Low Temperature}
One situation of interest is at low temperature when deep quantum tunneling is present.       
The tunneling probability decreases rapidly with temperature and the corresponding polymer tends to remain localized on one side of the barrier. 
When this happens the two end beads of the open polymer are not able to explore a sufficiently large interval of $x$ in an unbiased MD run. 
Consequently the time necessary to achieve good sampling of $\tilde{n}(x)$ becomes prohibitively long. 
This difficulty, however, can be overcome by enhanced sampling techniques developed over the last two decades, such as, e.g. metadynamics \cite{meta}, variationally enhanced sampling \cite{ves}, and forward flux sampling \cite{forward_flux} etc., if one has a good order parameter that captures the slow dynamical mode(s) in the MD.   

In the case of the directional momentum distribution, this slow mode is typically along the tunneling direction where the potential energy barrier is high compared to $k_BT$.      
If $BC$ is the tunneling direction, $x$ is a good order parameter kinetically, as it facilitates fluctuation of the end beads of $A$ to cross the potential energy barrier and reach the long tails of $\tilde{n}(x)$.      

In this paper, we adopt a recently proposed technique called variationally enhanced sampling (VES) \cite{ves}, and use $x$ as the order parameter.  
\subsubsection{Variationally Enhanced Sampling}
Here we briefly review the basics of VES. 
VES considers a functional of the bias potential, $V_b(x)$, of the order parameter, which, in our case, is the modified end-to-end displacement, $x$: 
\begin{equation}
  \Omega[V_b(x)] = \frac{1}{\beta}\log\int dx e^{-\beta(F(x) + V_b(x))} + \int dx \, p_t(x)V_b(x)
\end{equation}
where $F(x)$ is the free energy profile of the order parameter $x$. $p_t(x)$ is a preset target probability distribution which will be taken to be uniform in the interval spanning the range of possible physical values for this quantity.   
This functional follows from the variational principle \cite{sm} of the Legendre transform of the convex functional $F[V_b] = \log \int dx \exp(-\beta(F(x) + V_b(x)))$ by treating $V_b(x)$ and $p_t(x)$ as the Legendre conjugate fields. 
It can be shown \cite{ves} that $\Omega[V_b]$ is a convex functional and its minimizer satisfies the following equation 
\begin{equation}
  V_{b,\min}(x) = -F(x) - \frac{1}{\beta}\log p_t(x) + C
\end{equation}
where $C$ is an unimportant constant. 
Thus, once $V_{b,\min}(x)$ is found, $F(x)$ can be obtained immediately.   
To find $V_{b,\min}(x)$, we first represent $V_b(x)$ by a finite linear expansion of basis functions $G_k(x)$, such as plane waves or Chebyshev polynomials,     
\begin{equation}
  V_b(x) \approx V_{b,\bm\alpha}(x) = \sum_{k} \alpha_k G_k(x)
\end{equation}
The convex functional $\Omega[V_b]$ then becomes a convex function of $\bm\alpha$, the expansion coefficients of $V_b$, and it can be minimized by a Newton-type method using the gradients and Hessians that can be calculated with MD sampling.  
See minimization details in \cite{ves}. 
\section{Numerical Example: Malonaldehyde}
\label{sec:mal}
\begin{figure}[H]
  \centering
  \includegraphics[scale=0.6]{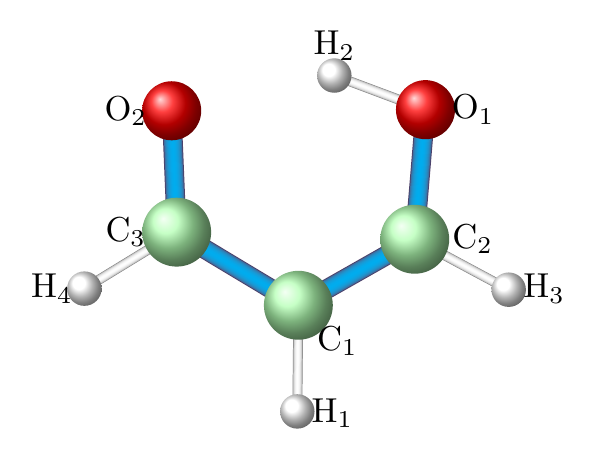}
  \caption{Malonaldehyde. 
  The distance between the O$_1$ atom and the O$_2$ atom in the lowest energy configuration is 4.87 a.u., according to the potential energy surface \cite{pes} used in this paper.}
  \label{fig:11}
\end{figure}

As a realistic example, we study the directional momentum distribution of Malonaldehyde (Fig. \ref{fig:11}). 
This molecule has been studied extensively experimentally \cite{mal_spectroscopy,tunneling_rotation,matrix_quenching,vibrational_tuning} and theoretically \cite{pes,tunneling_splitting} because features due to the tunneling hydrogen can be seen in its vibrational spectrum \cite{mal_spectroscopy}.  
Computational studies of the tunneling splitting with diffusion Quantum Monte Carlo \cite{pes} and PIMD \cite{tunneling_splitting} in Malonaldehyde show that the molecule is in the deep tunneling regime at inverse temperature $\beta = 5000$ a.u., i.e. at this temperature, the two lowest many-body energy eigenvalues dominate the energy spectrum. 
Here we are interested in whether we can obtain with PIMD the directional momentum distribution of the tunneling hydrogen atom (H$_2$) along the direction connecting the two oxygen atoms (O$_1$ and O$_2$). 
We also look for features in the directional momentum distribution when tunneling is present. 
The center of mass of the molecule in the configuration of lowest potential energy was chosen as the origin of the coordinates in the MD simulation. 

Our calculation used the potential energy surface of Malonaldehyde that was recently published \cite{pes}. 
Two VES calculations were performed. First, we froze all the atoms other than H$_2$ in the minimum energy configuration (Fig. \ref{fig:11}), and moved H$_2$ from $x = -2$ a.u. to 2 a.u. to obtain an effective one-dimensional (1D) potential energy profile. 
This 1D potential was then symmetrized about $x = 0$ to obtain the even potential energy profile $V_{\text{1D}}(x)$, shown in the top panel of Fig. \ref{fig:1d}. 
This potential was then used in an 1D PIMD calculation of the momentum distribution of the H$_2$ atom. 
$V_{\text{1D}}(x)$ was extended linearly outside the range $[-2, 2]$ a.u. to deal with the rare cases where an H$_2$ bead moved beyond this range.
In the second VES calculation we allowed all the atoms in the molecule to move freely in a many-body PIMD calculation, using the full many-body potential energy surface.   

\subsection{Simulation Details}
PIMD with a large bead number suffers ergodicity problem when using a standard Langevin thermostat \cite{}, as the frequency spectrum of the free polymer chain becomes broader as the number of beads increases.
To overcome this ergodicity problem, we adopt here a generalized Langevin equation (GLE) thermostat \cite{gle} designed to have an near-optimal relaxation time over a wide-frequency range to achieve a much better thermostating efficiency. 
The GLE matrices that we used are given in the supplementary material \cite{sm}. 
In the variational calculation, the first 12 even Chebyshev polynomials of the first kind, often referred to as the $T$-Chebyshev polynomials, were used as the basis functions, i.e. $\{T_2(x), T_4(x), \cdots, T_{24}(x)\}$ were used to expand $V_b(x)$.
The target distribution of $x$ was taken to be a uniform distribution between $-3.0$ a.u. and $3.0$ a.u. and zero outside this range. 
The displacement $x$ was forced to span an interval smaller than 6.0 a.u., by setting a reflective boundary for the beads at positions equal to $\pm$3.0 a.u. along $BC$. 
The widest allowed displacement of 6 a.u. should be compared with a distance of 4.87 a.u. between the two oxygens (O$_1$ and O$_2$) in the molecular configuration of lowest potential energy.   

In the 1D calculation, the inverse temperature was set at $\beta = 5000$ a.u..    
In the many-body calculation, inverse temperatures $\beta = 1000, 3000,$ and $5000$ a.u. were used.      
The center of mass position was kept fixed in the simulation by removing the center of mass velocity acquired from the thermostat at each step. 
The rest of the VES parameters are given  in the Table \ref{table:ves_parameter}. 
\begin{table}[htb!]
  \setlength{\tabcolsep}{1em}
  \begin{tabular}{l  l  l  l  l} 
    \hline
    \hline
    $\beta$ & MD steps & $\delta$t & $\mu$  & bead number\\
    \hline
    5000 & 12500 & 10 &  0.0001 & 400\\
    \hline
    \hline
  \end{tabular}

  \vspace{5mm} 
  \begin{tabular}{l  l  l  l  l} 
    \hline
    \hline
    $\beta$ & MD steps & $\delta$t & $\mu$  & bead number\\
    \hline
    1000 & 1200 & 5 &  0.0004 & 84\\
    \hline
    3000 & 1200 & 5 & 0.0001 & 84\\
    \hline
    5000 & 1000 & 10 & 0.0001 & 170\\
    \hline
    \hline
  \end{tabular}
  \caption{VES parameters in a.u.. The upper table is for the 1D calculation, while the lower table is for the many-body case. 
MD steps is the number of MD steps used for each variational step to sample the gradient and the Hessian of $\Omega$. $\delta t$ is the time step of the MD. 
$\mu$ is the step size of the gradient descent in the VES minimization of $\Omega$. (see Eq. 11 in the original paper \cite{ves} for the gradient descent step of the minimization.)
The calculation is done with 16 walkers in parallel to speed up sampling.  
}
\label{table:ves_parameter}
\end{table}
We checked for convergence with respect to the number of beads used in PIMD, finding that the converged number of beads agreed with the number used in Ref. \cite{tunneling_splitting} for the same system at the same temperature to study similar tunneling configurations.
\subsection{Results of the 1D Simulation}
\begin{figure}[]
  \centering
  \includegraphics[scale=0.40]{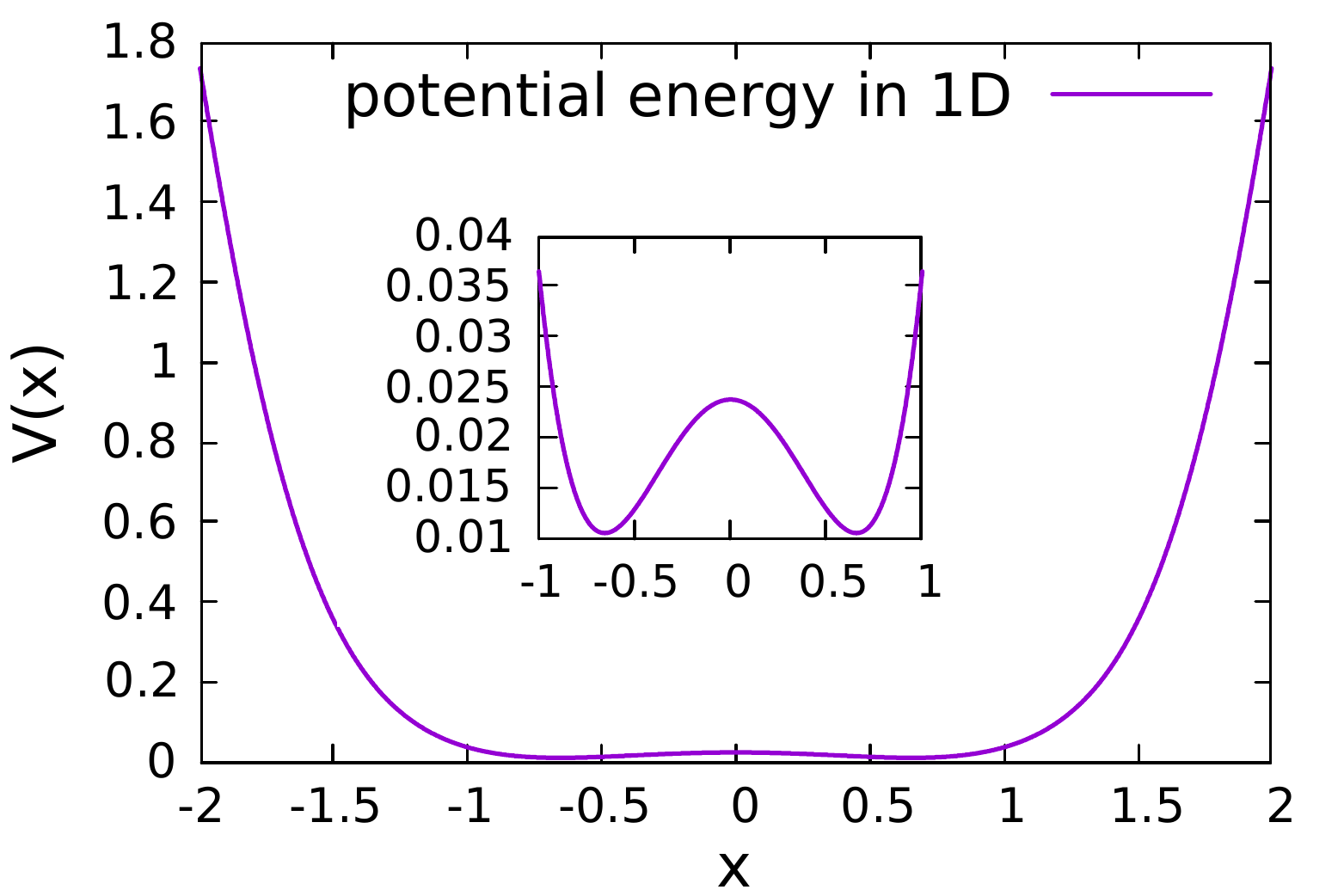}
  \includegraphics[scale=0.40]{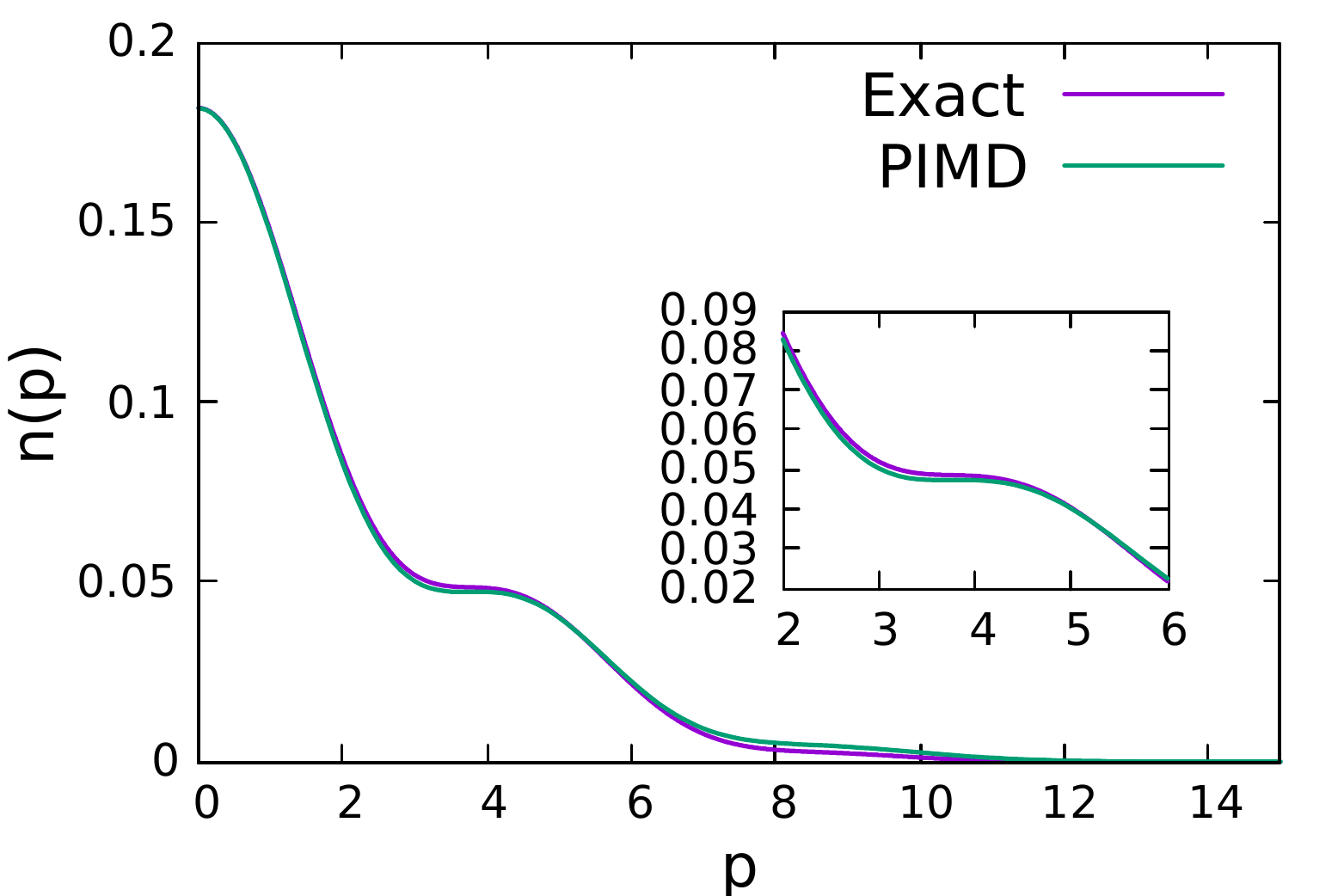}
  \caption{Top: The potential energy used in the 1D PIMD calculation. 
  Bottom: The momentum distribution of the H$_2$ atom obtained from the 1D PIMD simulation and solving the eigen-wavefunctions. 
Note the secondary feature represented by a prominent shoulder at approximately 4 a.u. in the distribution. In both panels the insets display details on a magnified scale. 
}
  \label{fig:1d}
\end{figure}

The 1D simulation was done to check whether a secondary feature exists in the momentum distribution of the tunneling particle in one dimension.      
Fig. \ref{fig:1d} shows that it does for the present 1D model potential.
In addition to the PIMD calculation, the momentum distribution was also obtained from numerically solving the 1D Schrodinger equation, yielding essentially the exact distribution.  
The two approaches agree very well, especially considering the sampling difficulty posed by the low temperature.    
In addition to the statistical error, the residual deviation between the PIMD simulation and the exact solution can be due to the truncation error in the basis functions and the finite number of PIMD beads.     
\subsection{Results of the Many-body Simulation}
\subsubsection{Convergence of VES}
We check the convergence of VES in two ways. 
One check consists in looking at the evolution of the variational parameters with respect to the variational step. 
The other check is to look at the distribution of the modified end-to-end displacements, $x$, under the minimizing bias potential.  
\begin{figure}[]
  \centering
  \includegraphics[scale=0.40]{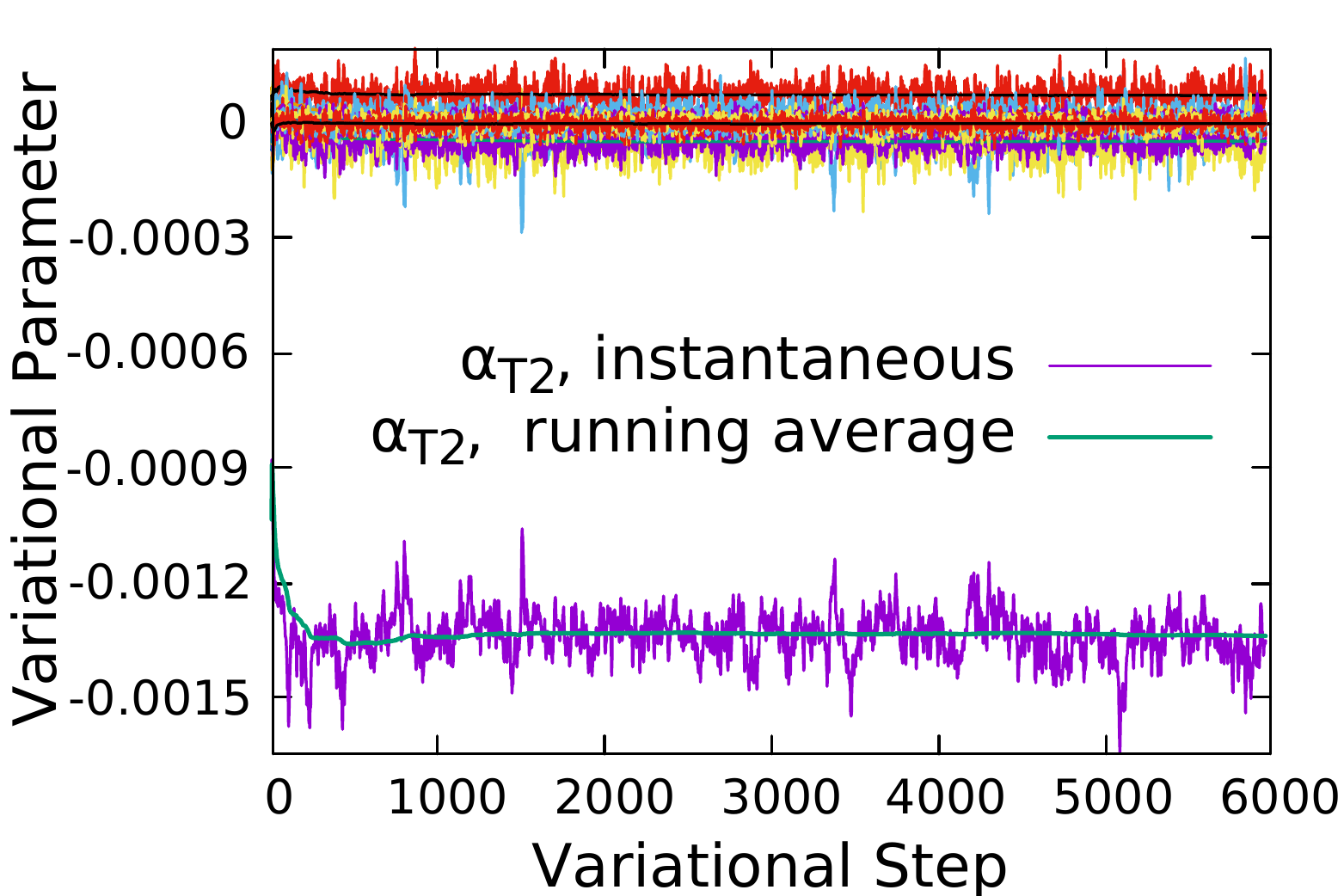}
  \includegraphics[scale=0.40]{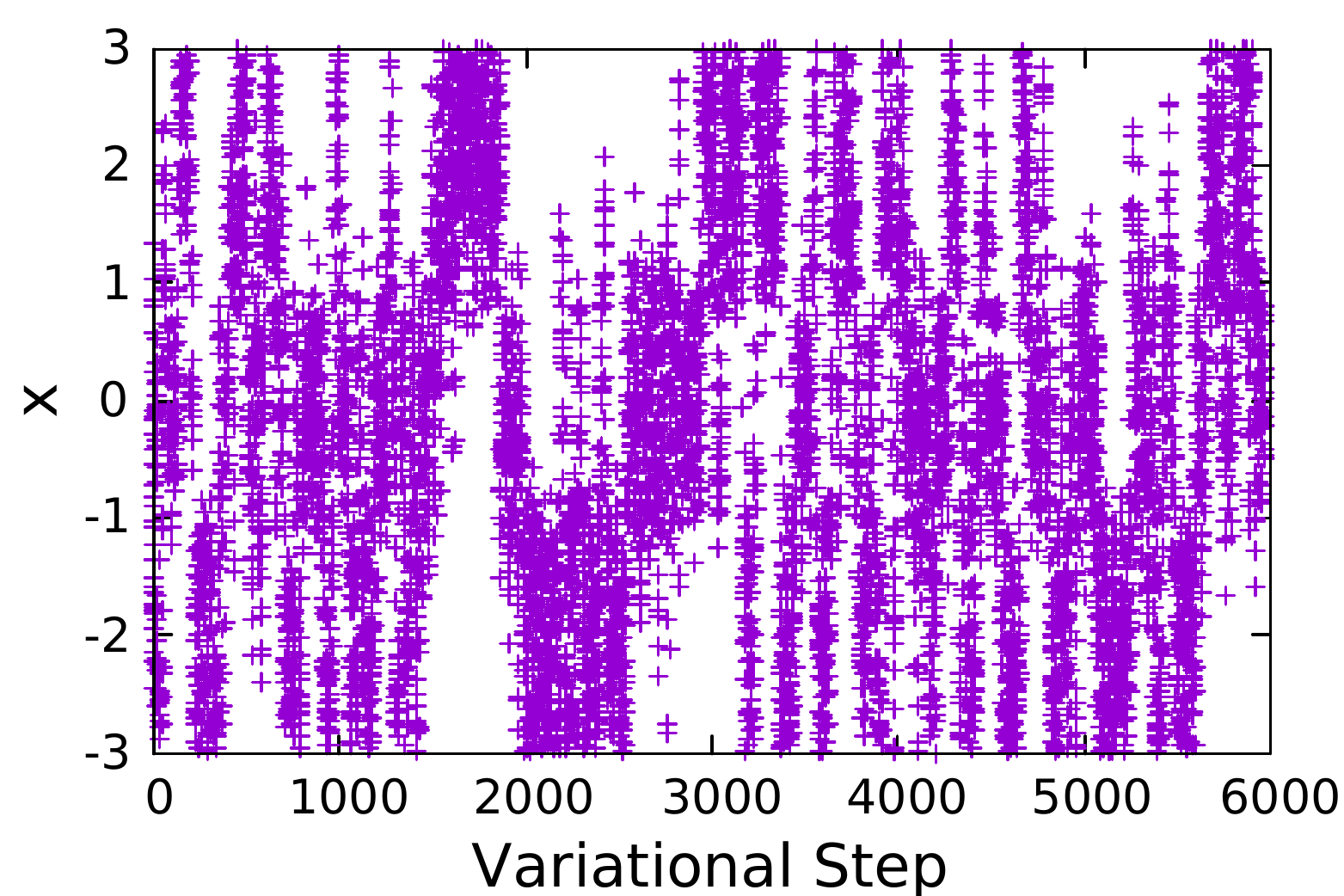}
  \caption{Result in a.u. of a VES calculation at $\beta=5000$ a.u.. 
  Top: The evolution of variational parameters. The coefficient for the basis function $T_2$ Chebyshev polynomial is labelled by $\alpha_{T2}$. 
The coefficients of the other basis functions, which appear as the data points around zero, are much smaller than that of $T_2$, and are thus not labelled.  
Bottom: The distribution of $x$ after quasi-stationarity of the variational coefficients has been reached.}
  \label{fig:x_5000}
\end{figure}
Fig. \ref{fig:x_5000} (top) shows the convergence of  the variational parameters in the $\beta=5000$ a.u. calculation starting with initial variational parameters taken from the result of a $\beta=3000$ a.u. VES calculation. 
Fig. \ref{fig:x_5000} (bottom) displays the distribution of $x$ during the variational simulation.  
We do see a uniformly fluctuating distribution of $x$, as required by the target distribution. 
Thus, $x$ explores all the available range without being trapped in a local potential energy minimum, indicating the occurrence of tunneling configurations in the simulation. 
We also checked that in an unbiased sampling at $\beta = 5000$ a.u., the order parameter $x$ is confined to the range $[-1, 1]$ and the system rarely tunnels within the computational time of the simulation.  
\subsubsection{$\tilde{n}(x)$ and $n(p)$}
\begin{figure}[]
  \centering
  \includegraphics[scale=0.40]{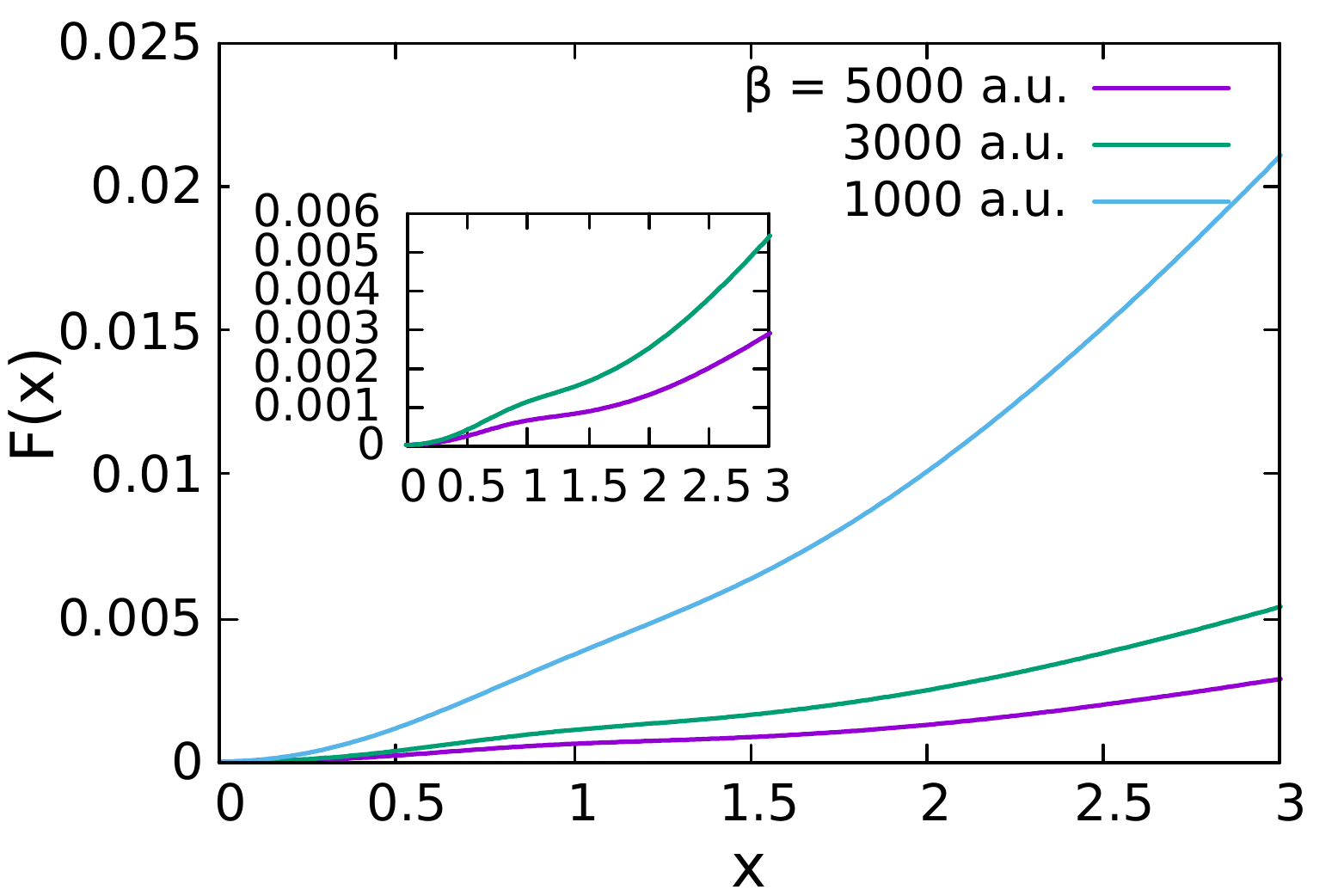}
  \includegraphics[scale=0.40]{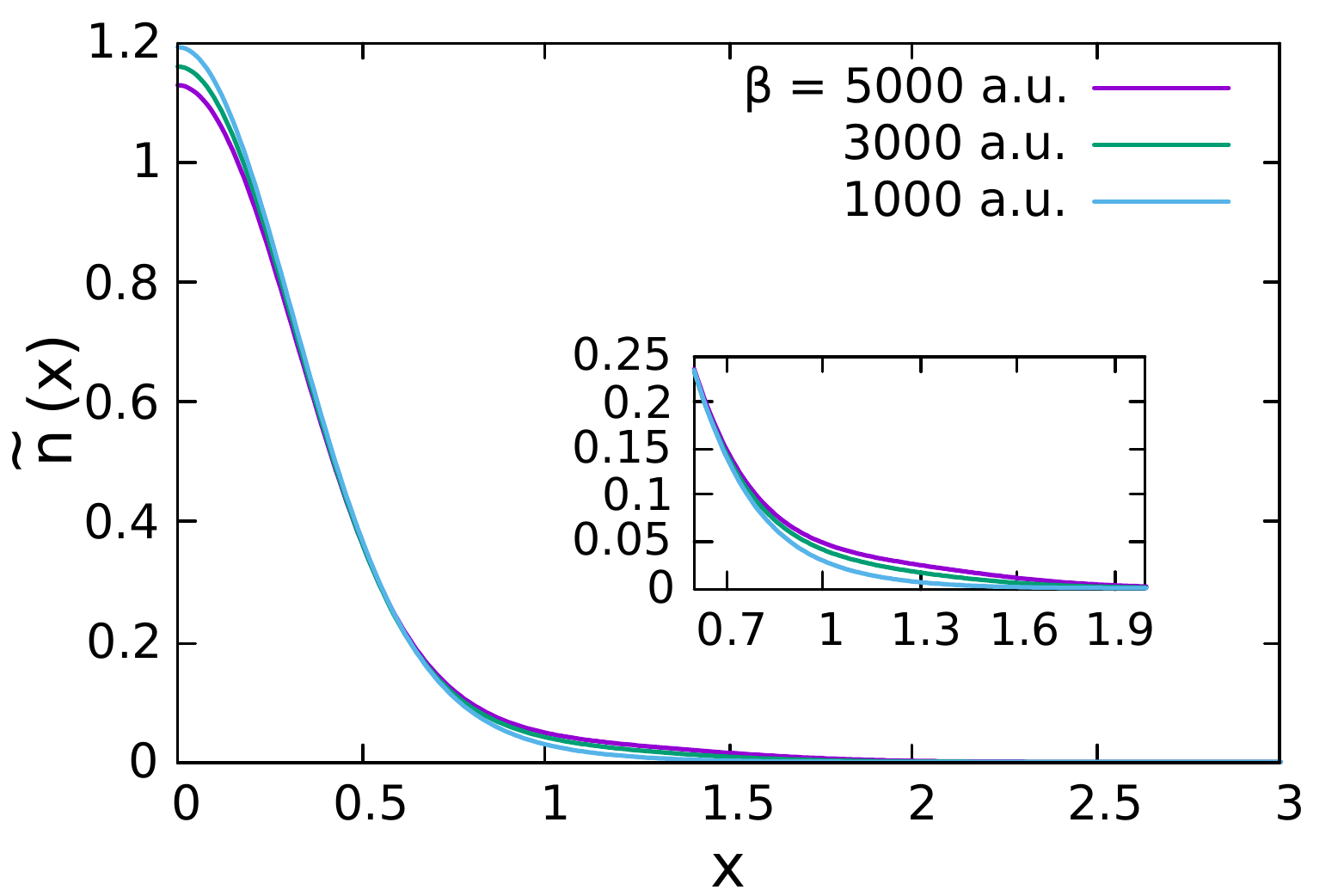}
  \includegraphics[scale=0.40]{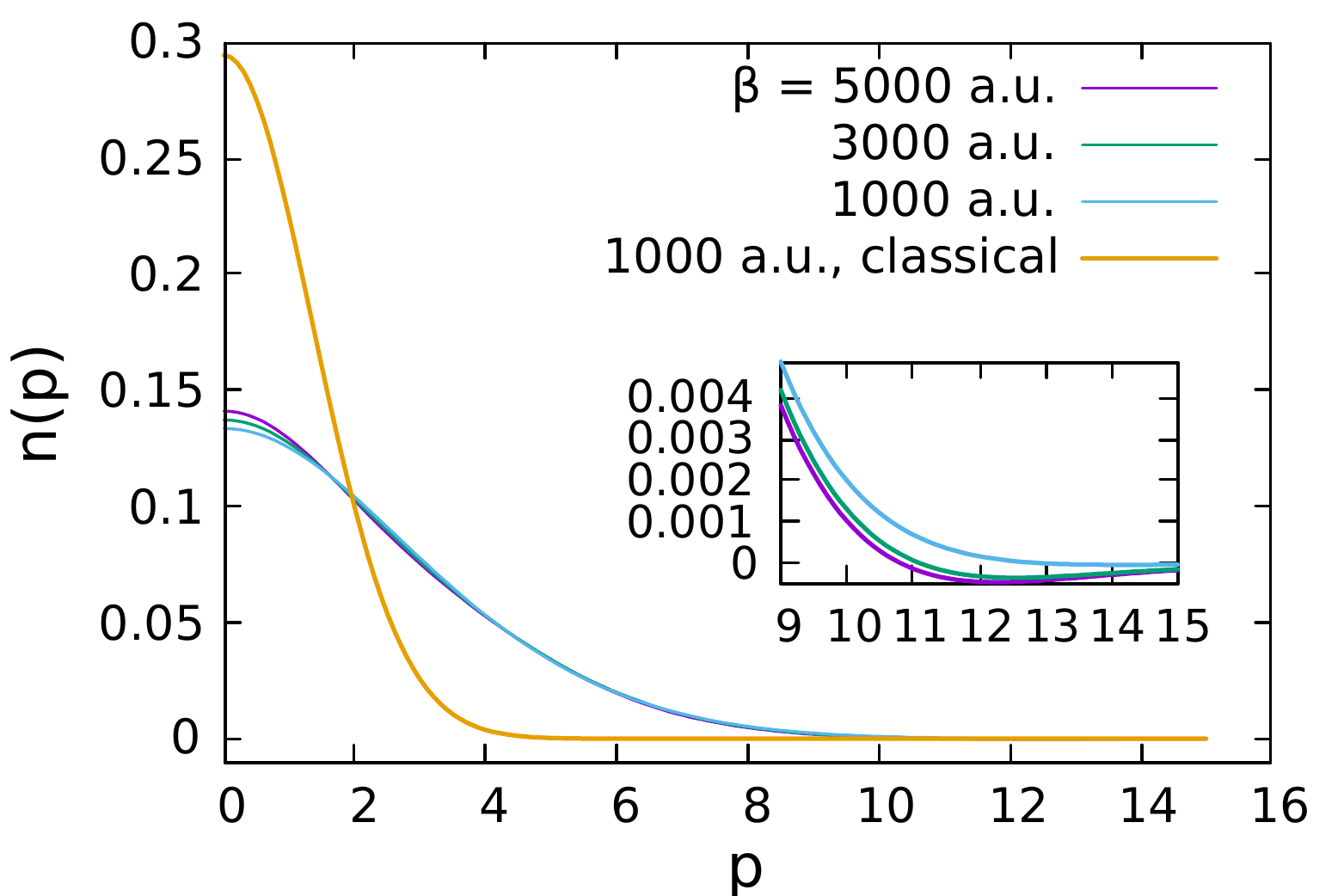}
  \caption{
  Top: free energy profile of $x$ with an inset showing $\beta = 3000$ and $5000$ a.u..
  Middle: the distribution of $x$ with an inset showing $\tilde{n}(x)$ for $x = 0.6$ to $2$. 
  Bottom: The directional momentum distribution with an inset showing $n(p)$ for $p = 9$ to 15.}
  \label{fig:x}
\end{figure}

In Fig. \ref{fig:x}, we present the free energy profile of $x$, $F(x)$, the directional end-to-end distance distribution, $\tilde{n}(x) \propto e^{-\beta F(x)}$, and the directional momentum distribution, $n(p) = \mathcal{F}\{\tilde{n}(x)\}$, where $\mathcal{F}$ denotes the Fourier transform, for $\beta = 1000, 3000,$ and $5000$ a.u..  
For reference, we also plot the momentum distribution from classical statistical mechanics. 

We use statistical bootstrap \cite{Bootstrap} to obtain the statistical error of the distributions. 
To obtain independence of the variational coefficients, for each $\alpha_k$, we use the block averaging method to obtain the block size by which, when grouped, the variational parameters become effectively uncorrelated in variational time. 

The bootstrap method is performed for $\tilde{n}(x)$ and $n(p)$.  
Bootstrap re-sampling done 100 times was found to give convergent results on the standard deviation of $\tilde{n}(x)$ and $n(p)$ for given $x$ and $p$ respectively.  
Uncertainties at selected $x$ and $p$ are tabulated in Table. \ref{table:uncertainty}.   
\begin{table}[htb!]
  \setlength{\tabcolsep}{0.7em}
  \begin{tabular}{l  l  l  l  l  l} 
    \hline
    \hline
    $\beta$ & vsteps & $\delta\tilde{n}_x(0)$ &  $\delta \tilde{n}_x(1)$ & $\delta n_p(0)$ &$\delta n_p(12)$ \\
    \hline
    1000 & 4328 & 0.005 & 0.0004 &  0.0005 & 0.00005\\
    \hline
    3000 & 4328 & 0.01 & 0.001 & 0.0013 & 0.0001\\
    \hline
    5000 & 5964 & 0.01 & 0.0013 & 0.0013 & 0.0001\\
    \hline
    \hline
  \end{tabular}
  \caption{Statistical uncertainty on $\tilde{n}(x)$ and $n(p)$. Here vsteps is the length of the variational trajectory at quasi-stationarity, $x=1$ a.u. is approximately where the plateau in $F(x)$ is, and $p=12$ a.u. is approximately where the first minimum (Fig. \ref{fig:x} inset) of $n(p)$ is.}
  \label{table:uncertainty}
\end{table}
\subsection{Discussion}
The quantum character of the distribution is most clearly seen in the comparison between the classical Boltzmann momentum distribution and the distribution sampled by PIMD.      
The momentum distribution is strongly broadened by the quantum effect. 
At this deep tunneling regime, the difference among the quantum momentum distribution across $\beta = 1000 - 5000$ a.u. is not nearly as close as that between the classical and quantum difference, indicating that the distribution is dominated by quantum, instead of thermal, fluctuations.       
One does see a deviation from the Gaussian behavior of $n(p)$, most pronounced in the free energy profile in Fig. \ref{fig:x}, as $F(x)$ is clearly different from a quadratic function of $x$, especially at low temperature.      
The inset of Fig. \ref{fig:x} shows a very shallow local minimum in $n(p)$, unfortunately with a small non-physical negative value.  
Although $\tilde{n}(x)$ is guaranteed to be everywhere positive by the requirement that $\tilde{n}(x) \propto \exp(-\beta F(x))$, there is no guarantee that, $n(p)$, the Fourier transform of $\tilde{n}(x)$, will be everywhere positive, and any statistical uncertainty in the results can lead to negative values of $n(p)$.    
{
In fact, the shallow minimum of the many-body $n(p)$ happens at around $p=$ 12 a.u., which in the 1D calculation is approximately the onset of the near-zero exponential tail of the momentum distribution.     
}
The secondary feature of $n(p)$ that is associated with ground-state tunneling in one-dimensional potentials is not observed in our many-body simulations beyond statistical uncertainty. 
\section{Sampling of reduced density matrix}
\label{sec:rdm}
To investigate further the reason for the absence of the secondary feature in $\tilde{n}(x)$ and $n(p)$, we study the reduced density matrix $\tilde{\rho}(r, r')$, symmetric in $r$ and $r'$, associated with the directional momentum distribution.
It is defined by requiring that $\tilde{n}(x)$ be related to it in the same way as in a strict 1D case:  
\begin{equation}
  \tilde{n}(x) \equiv \int dr' dr \tilde{\rho}(r, r')\delta(r-r'-x)
\end{equation}
In the context of our PIMD calculation, a natural definition is to take $\tilde{\rho}(r,r')$ to be the probability distribution of the order parameter $r(\vec r(\tau), x)$ and $r'(\vec r(\tau), x)$,   
\begin{equation}
  r(\vec r(\tau), x) \equiv \vec r_A(0)\cdot \frac{\vec r_B(0) - \vec r_C(0)}{\abs{\vec r_B(0) - \vec r_C(0)}} 
\end{equation}
and 
\begin{equation}
  r'(\vec r(\tau), x) \equiv \vec r_A(\beta \hbar)\cdot \frac{\vec r_B(0) - \vec r_C(0)}{\abs{\vec r_B(0) - \vec r_C(0)}} 
\end{equation}
Then, from Eq. \ref{eq:nx_PI}, $\tilde{\rho}(r,r')$ can be defined as  
\begin{equation}
  \label{eq:rho_1D}
  \begin{split}
  \tilde{\rho}(r, r') \propto &\int dx \int \mathcal{D}[\vec r_{\not=A}(\tau)] \int \mathcal{D}[\vec r_{A}(\tau)] \\ 
  &\delta(r - r(\vec r(\tau), x)) \,\, \delta(r'-r'(\vec r(\tau), x))\,\, e^{-\frac{S[\vec r(\tau)]}{\hbar}}
\end{split}
\end{equation}
with the same boundary condition on $\vec r(\tau)$ and the same action as in Eq. \ref{eq:nx_PI}.  
Making the change of variable in Eq. \ref{eq:r_tilde}, the same Hamiltonian of Eq. \ref{eq:hamiltonian} can be used to sample $\tilde{\rho}(r,r')$. 
Again to overcome the difficulty in sampling the two-dimensional order parameter $(r', r)$, we use VES to facilitate the simulation.

The target distribution is taken to be the uniform distribution within the square domain in which each of the two variables of the order parameter $(r',r)$ is restricted to the interval $[-1.8, 1.8]$ in a.u. by a reflective boundary wall at the boundary of the domain. 
The basis functions are taken to be the product basis of the first 11 $T$-Chebyshev polynomials, i.e. $G_{ij}(r, r') = T_i(r) T_j(r')$ for $i, j = 0, 1, 2, 3, \cdots, 10$. 
That is, a total of 121 basis functions are used to represent the free energy profile of $(r,r')$.  
Again, the reduced density matrix is sampled for both the one-dimensional and the many-body system as in the calculation of the directional momentum distribution.  
The calculation is performed at inverse temperature $\beta = 5000$ a.u.. 
The other simulation parameters are the same as in the calculation for $\tilde{n}(x)$, except that in this case 5000 MD steps are used for the many body calculation.  
\subsection{Results}
\label{subsec:rdm_result}
We first check that the directional momentum distribution can indeed be reproduced with the reduced density matrix.  
After this check, $\tilde{\rho}(r, r')$ is discretized to compute its spectrum, which is listed in Table \ref{table:spec}.     
\begin{table}[htb!]
  \centering
  \begin{tabular}{lrr} 
    \hline
    \hline
     1D Exact & 1D PIMD  & Many-body PIMD\\
     \hline
     0.67700   & 0.675(5) & 0.454(1)\\ 
     0.32300   & 0.326(4) & 0.394(2)\\ 
    $10^{-18}$ & -0.002(1) & 0.074(1)\\ 
    $10^{-24}$ & 0.001(1) & 0.051(1) \\ 
    $10^{-36}$ & -0.001(1) & 0.0068(3)\\ 
    \hline
    \hline
  \end{tabular}
  \caption{The first five eigenvalues of $\tilde{\rho}(r, r')$ at $\beta = 5000$ a.u.. 
The exact spectrum (shown in the first column) is also obtained by solving the Schrodinger Equation in the one-dimensional model for comparison. 
The number in the parenthesis is the uncertainty on the last digit.
In the 1D model, the eigenvalues other than the first two are prohibitively small and an exact determination of them is beyond the accuracy of the PIMD simulation, so that any statistical uncertainty could lead to an unphysical negative value.      
}

  \label{table:spec}
\end{table}

\begin{figure}[]
  \centering
  \includegraphics[scale=0.4]{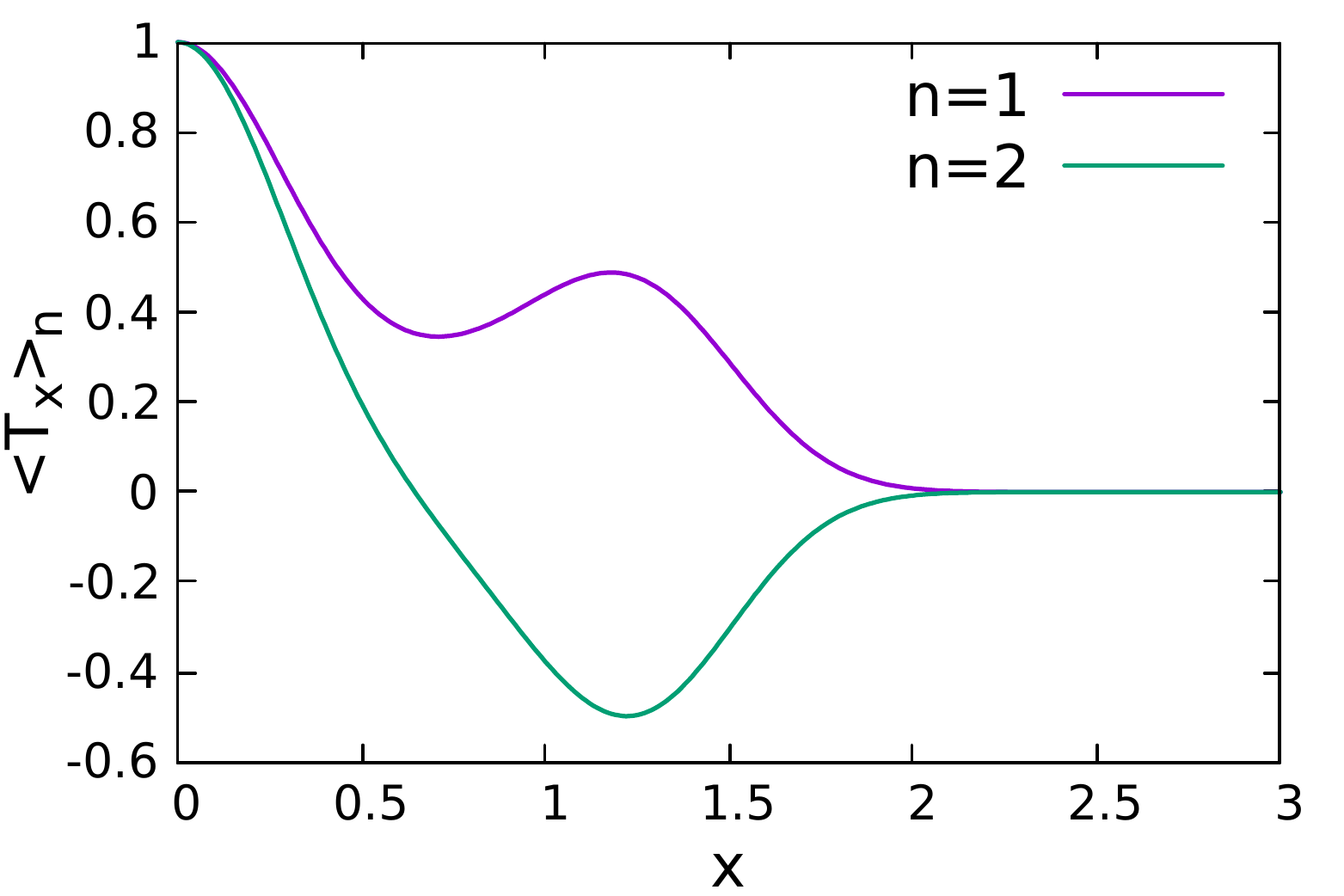}
  \includegraphics[scale=0.4]{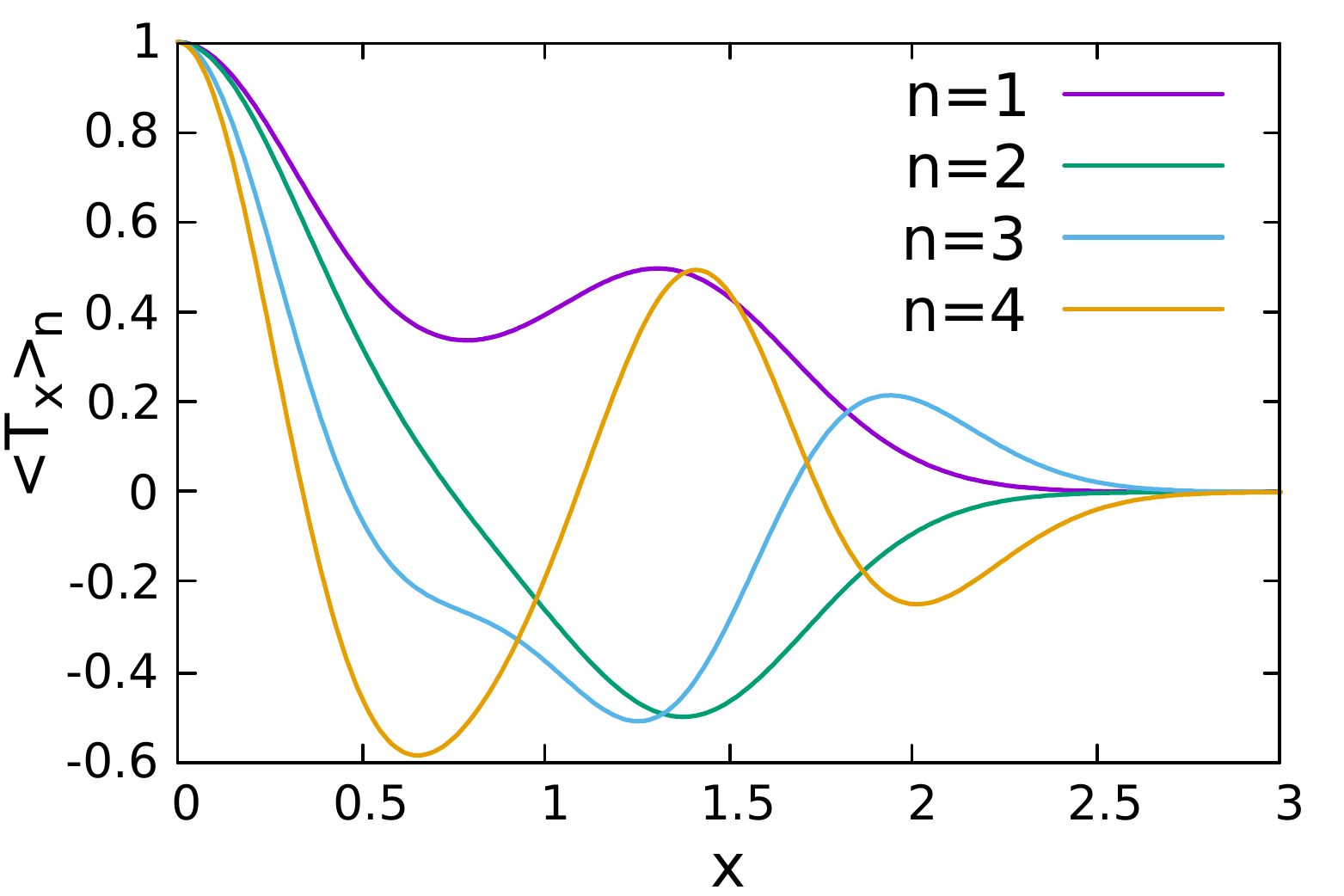}
\caption{Top: $\braket{\hat{T}_x}_n$ for the first two eigenstates of $\tilde{\rho}$ in the one-dimensional system. 
Bottom: $\braket{\hat{T}_x}_n$ for the first four eigenstates of $\tilde{\rho}$ in the many-body case. 
}
  \label{fig:Tx}
\end{figure}

In the representation of the eigenstates of the reduced density matrix, the distribution of the end-to-end distance can be calculated as in the following (see \ref{sec:translation} for a proof),      
\begin{equation}
  \label{eq:translation}
  \tilde{n}(x) = \sum_{n=1}\rho_n \int dr dr' \psi_n^*(r') \psi_n(r + x) = \sum_{n=1} \rho_n \braket{\hat{T}_x}_n
\end{equation}
where $\rho_n$ is the $n$th eigenvalue of the reduced density matrix, $\psi_n$ is the corresponding eigenstate, $\hat{T}_x$ is the translation operator for a displacement $x$, and $\braket{\hat{T}_x}_n$ is its expectation value in the $n$th eigenstate.   
Viewed as a function of $x$, each $\braket{\hat{T}_x}_n$ has its own distinct features shown in Fig. \ref{fig:Tx}, such as secondary peaks and valleys.   
However, as they are superimposed as a weighted sum to produce $\tilde{n}(x)$, features associated with each $n$ tend to cancel each other.   
If, however, the ground state dominates the density matrix, for example, in the case of ground state tunneling, then features of $\braket{\hat{T}_x}_1$ survive into $\tilde{n}(x)$, and the secondary feature in $n(p)$ will be present, as in the case of the one-dimensional model.        

In the one-dimensional model, the reduced density matrix is dominated by the first two eigenstates with the ground state having a definitively larger weight.     
Thus, despite the partial cancellation of $\braket{\hat{T}_x}_1$ by $\braket{\hat{T}_x}_2$, a secondary shoulder is still present in $\tilde{n}(x)$.   
As the temperature is lowered even more, the secondary feature of $\tilde{n}(x)$ is even more pronounced.  

In the many-body case, however, the situation is more complicated.  
At inverse temperature $\beta = 5000$ a.u., the Malonaldehyde molecule is in the deep tunneling regime, meaning that only its first two energy eigenstates contribute significantly to the full density matrix \cite{tunneling_splitting}.  
The tunneling splitting energy of this molecule has been determined to be $\Delta E = 20 \pm 1$ cm$^{-1}$ by both diffusion Quantum Monte Carlo \cite{pes} and PIMD \cite{tunneling_splitting}. 
This means that at $\beta = 5000$ a.u., the weight of the ground state in the full density matrix is $\frac{1}{1 + \exp(-\beta \Delta E)} = 61\%$, which is rather close to the 67\% found in the 1D case.  
Thus, one might naively expect that a secondary feature should be present in the momentum distribution. 
The first eigenstate of the reduced density matrix in the many-body case, however, only contributes 45\% of the trace, and the first two states only 82\%, leaving a nontrivial weight for the higher-lying states, indicating significant quantum entanglement.       
Although each $\braket{\hat{T}_x}_n$ of the many-body system is no less featured than that in the 1D system, the secondary features of $\braket{\hat{T}_x}_1$ are canceled by the higher eigenstates of $\tilde{\rho}$ to a much larger extent, and do not persist into $\tilde{n}(x)$.          
In addition, unlike the case in the one-dimensional model where lowering the temperature enhances the secondary feature of the momentum distribution by eventually populating only the ground state, the directional momentum distribution may never exhibit a secondary feature no matter how low one pushes the temperature to be, because of the fundamental limitation posed by the quantum entanglement.    
\subsection{Extrapolation to zero temperature}
\begin{figure}[]
  \centering
  \includegraphics[scale=0.4]{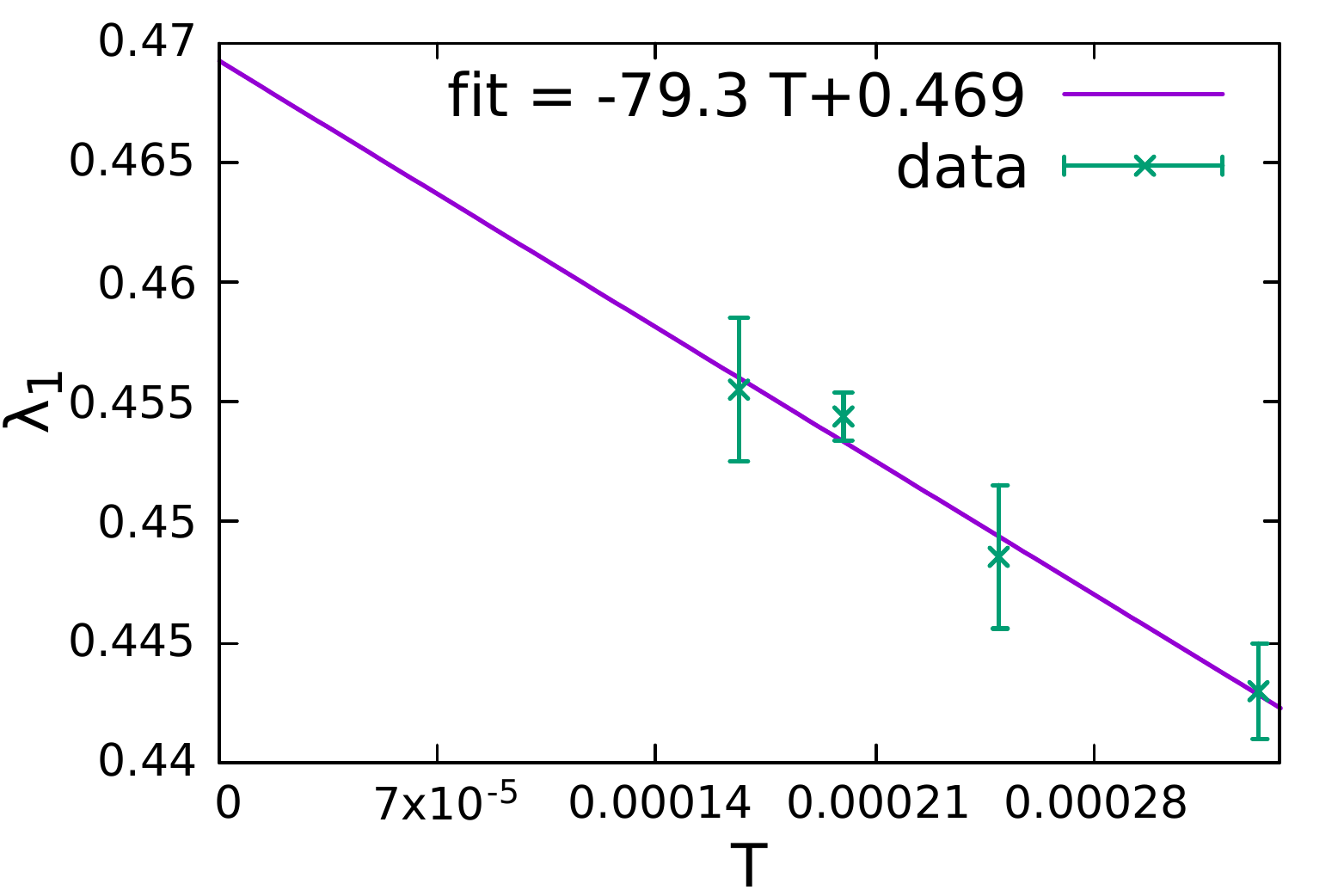}
  \includegraphics[scale=0.4]{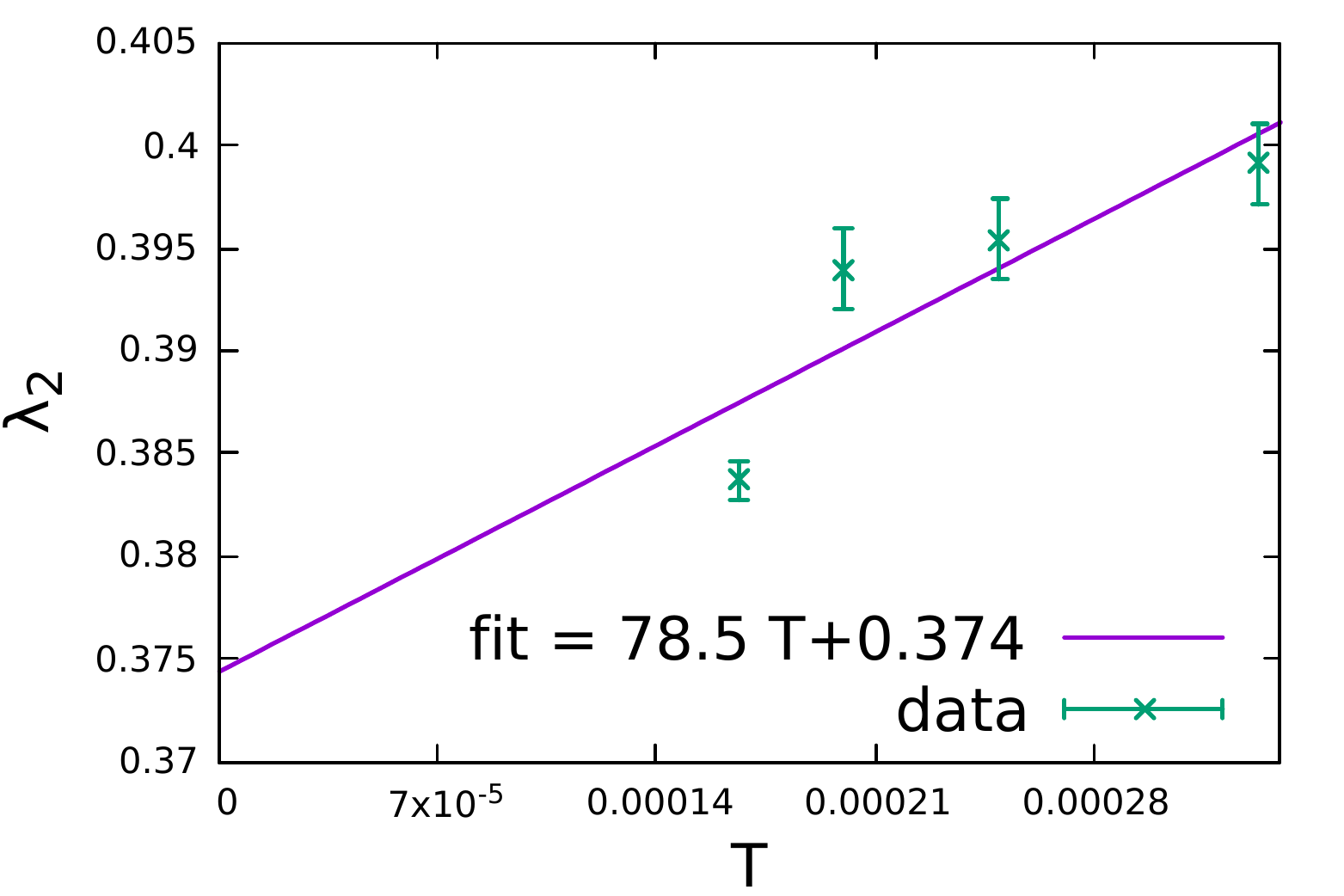}
\caption{Top: The extrapolation to zero temperature for the leading eigenvalue of the reduced density matrix. 
Bottom: The extrapolation to zero temperature for the sub-leading eigenvalue of the reduced density matrix. 
}
  \label{fig:fit}
\end{figure}

Additional evidence of the quantum entanglement can be obtained by extrapolating the eigenvalues of the reduced density matrix to zero temperature.  
We computed the leading and subleading eigenvalues, $\lambda_1$ and $\lambda_2$, of the reduced density matrix $\rho(r,r')$ for $\beta = 3000, 4000, 5000,$ and $6000$ a.u. using the procedure above.  
The result is given in Fig. \ref{fig:fit}. 
As the system is finite, we do not expect any non-analytic temperature-dependence of $\lambda_1$ and $\lambda_2$, and perform a linear extrapolation to zero temperature. 
We obtained $\lambda_1(T = 0) = 0.469$ and $\lambda_2(T= 0) = 0.374$, suggesting significant entanglement even at the zero temperature. 


\section{Summary}
\label{sec:summary}
In this paper we have derived a proper PIMD estimator for the directional momentum distribution of a quantum particle, where the projection is defined with reference to the internal coordinates of the atomistic system.    
This distribution reduces the three-dimensional momentum distribution of a particle to one dimension, serving as a particularly suitable comparison to one-dimensional systems.  
In addition, this reduction allows the sampling of the directional momentum to depend only on the internal motion of the system, which is much faster than the overall rotation of the system.      
At the deep tunneling regime of a quantum particle, where the unbiased sampling is difficult, enhanced sampling techniques, such as VES, make the sampling possible. 
In the example molecule of Malonaldehyde, we find that the secondary features in $n(p)$ for one-dimensional double-well potentials are not present in the many-body system beyond statistical uncertainty, due to the presence of quantum entanglement.    

The directional momentum distribution may be studied in other systems in the future.   
For example, it has been suggested \cite{collective_tunneling} that in the ice-6 phase of water, the hydrogen atoms tunnel concertedly around a hexagonal ring formed by the oxygen atoms.
In this example, the directional momentum seems particularly fitting to study the correlation in the proton tunneling along directions defined by the positions of the oxygen atoms.      

In addition, the modified momentum distribution is not limited to longitudinal momentum.   
For example, one may consider the distribution of transverse momentum, $\hat{\vec p}_A \times \frac{\hat{\vec r}_B - \hat{\vec r}_C}{\abs{\hat{\vec r}_B - \hat{\vec r}_C}}$,   
by similar techniques in other cases of interest. 

\section{Supplementary Material}
See supplementary matetrial for the GLE matrices used to do the PIMD sampling. 
\section{Acknowledgements}
We gratefully acknowledge support from the DOE Award DE-SC0017865.
\section{Appendix}
\subsection{Estimator of the directional momentum distribution}
\label{sec:dist_derivation}
The directional momentum distribution is equal to the quantum statistical average of the directional momentum distribution operator,  
\begin{widetext}
\begin{equation}
\label{eq:np}
\begin{split}
  n(p) &= \Tr (\delta(\hat{\vec p}_A \cdot \frac{\hat{\vec r}_B - \hat{\vec r}_C}{\abs{\hat{\vec r}_B - \hat{\vec r}_C}} - p)\hat{\rho})\\
  &= \int d^{3N}\vec r d^{3N}\vec r' \braket{\vec r | \delta(\hat{\vec p}_A \cdot \frac{\hat{\vec r}_B - \hat{\vec r}_C}{\abs{\hat{\vec r}_B - \hat{\vec r}_C}} - p) | \vec r'} \braket{\vec r'|\hat{\rho}|\vec r}   \\ 
  &= \int d^{3N}\vec r d^{3N}\vec r' \braket{\vec r | \delta(\frac{\hat{\vec r}_B - \hat{\vec r}_C}{\abs{\hat{\vec r}_B - \hat{\vec r}_C}} \cdot \hat{\vec p}_A - p) \left(\int d^{3N}\vec p' \ket{\vec p'}\bra{\vec p'}\right)|\vec r'} \braket{\vec r'|\hat{\rho}|\vec r}   \\ 
  &= \int d^{3N}\vec r d^{3N}\vec r' \int d^{3N} \vec p' \braket{\vec r | \delta(\frac{\vec r_B - \vec r_C}{\abs{\vec r_B - \vec r_C}} \cdot \vec p'_A - p) \ket{\vec p'}\bra{\vec p'}\vec r'} \braket{\vec r'|\hat{\rho}|\vec r}    
\end{split}
\end{equation}
where $\vec r_B$ and $\vec r_C$ are three dimensional vectors which make up parts of the $3N$-dimension vector $\vec r = \vec r_A \otimes \vec r_B \otimes \vec r_C \otimes ...$. Similarly, $\vec p'_A$ is a 3D vector which is the part associated with atom $A$ of the $3N$-dimension vector $\vec p'$.
  \begin{equation}
  \begin{split}
&= \int d^{3N} \vec r d^{3N} \vec r'  \int d^3 \vec p'_A \delta(\vec p'_A \cdot \frac{\vec r_B - \vec r_C}{\abs{\vec r_B - \vec r_C}} - p) \braket{\vec r_A |\vec p'_A} \braket{\vec p'_A | \vec r'_A}
\left(\int d^{3N-3} \vec p'_{\not= A} \braket{\vec r_{\not= A} | \vec p'_{\not= A}} \braket{\vec p'_{\not= A} | \vec r'_{\not= A}}\right)  \cdot \rho(\vec r', \vec r)\\
&= \int d^{3N} \vec r d^{3N} \vec r' \delta(\vec r_{\not=A} - \vec r'_{\not=A}) \int d^3 \vec p'_A \left(\frac{1}{2\pi\hbar}\right)^3 e^{\frac{i}{\hbar} \vec p'_A \cdot (\vec r_A - \vec r'_A)} \delta(\vec p'_A \cdot \frac{\vec r_B - \vec r_C}{\abs{\vec r_B - \vec r_C}} - p) \cdot \rho(\vec r', \vec r)
\end{split}
\end{equation}
We then use the mathematical identity (see Sec. \ref{sec:delta} for a proof) 
\begin{equation}
  \label{eq:delta_integral}
  \int d^3\vec p e^{i\vec p \cdot \vec a} \delta(\vec p \cdot \vec b - p)  = (2\pi)^2 \int dx \, e^{ipx} \, \delta(-\vec a + x \vec b) 
\end{equation} 
to write Eq. \ref{eq:np} as
\begin{equation}
  n(p) = \frac{1}{2\pi\hbar}\int\, dx\, e^{ipx} \int d^{3N}\vec r d^{3N}\vec r' \delta(\vec r_{\not=A} - \vec r'_{\not=A}) \delta\left(\vec r'_A - \vec r_A + x \cdot \frac{\vec r_B - \vec r_C}{\abs{\vec r_B - \vec r_C}}\right) \rho(\vec r', \vec r) \equiv \frac{1}{2\pi\hbar} \int\, dx\, e^{ipx}\tilde{n}(x)
\end{equation}
which defines the modified end-to-end displacement $x$, and its distribution $\tilde{n}(x)$. 
\end{widetext}

\newpage
\subsection{Proof of Eq. \ref{eq:delta_integral}}
\label{sec:delta}
\begin{equation}
  \begin{split}
    \int d^3 \vec p e^{i\vec p \cdot \vec a} & \delta(\vec p \cdot \vec b - p) = \int d^3 \vec p e^{i\vec p \cdot \vec a} \frac{1}{2\pi} \int dx e^{-ix(\vec p \cdot \vec b - p)} \\
    &= \frac{1}{2\pi}  \int dx e^{ipx} \int d^3 \vec p e^{i\vec p \cdot \vec a} e^{-ix \vec p \cdot \vec b} \\
    &= (2\pi)^2 \int dx e^{ipx} \delta^{(3)}(-\vec a + x \vec b)
  \end{split}
\end{equation}
\subsection{Proof of Eq. \ref{eq:n(x)}}
\label{sec:transform}
First note that the boundary condition on $\vec r(\tau)$ is 
\begin{equation}
 \vec r_A(\beta\hbar) = \vec r_A(0) - x \cdot \frac{\vec r_B(0) - \vec r_C(0)}{\abs{\vec r_B(0) - \vec r_C(0)}}   
\end{equation}
After the substitution 
\begin{equation}
  \vec r_A(\tau) = \tilde{\vec r}_A(\tau) - y(\tau) \cdot x \cdot \frac{\vec r_B(0) - \vec r_C(0)}{\abs{\vec r_B(0) - \vec r_C(0)}}  
\end{equation}
with $y(\tau) = \frac{\tau}{\beta\hbar} - \frac{1}{2}$, the boundary condition of $\tilde{\vec r}(\tau)$ is 
\begin{equation}
  \tilde{\vec r}(\beta\hbar) = \tilde{\vec r}(0)
\end{equation}
To prove Eq. \ref{eq:n(x)}, one only needs do the following expansion  
\begin{align*}
  &\int_0^{\beta\hbar} \dot{\vec r}_A^2 d\tau = \int_0^{\beta\hbar} \left(\dot{\tilde{\vec r}}_A(\tau)- \dot{y}(\tau) \cdot x \cdot \frac{\vec r_B(0) - \vec r_C(0)}{\abs{\vec r_B(0) - \vec r_C(0)}} \right)^2 d\tau 
  \\
  &= \int_0^{\beta\hbar} (\dot{\tilde{\vec r}}_A^2 + \dot{y}^2 x^2)  d\tau - \int_0^{\beta\hbar}\dot{\tilde{\vec r}}_A  d\tau \cdot \frac{2x  (\vec r_B(0) - \vec r_C(0))}{\beta\hbar  \abs{\vec r_B(0) - \vec r_C(0)}}  
\end{align*}
Note that $\int_0^{\beta\hbar} \dot{\tilde{\vec r}}_A = \tilde{\vec r}(\beta\hbar) - \tilde{\vec r}(0) = 0$, thus 
\begin{equation}
  \int_0^{\beta\hbar} \dot{\vec r}_A^2 d\tau = \int_0^{\beta\hbar} (\dot{\tilde{\vec r}}_A^2 + \dot{y}^2 x^2)  d\tau = \int_{0}^{\beta\hbar} \dot{\tilde{\vec r}}_A^2 d\tau + \frac{x^2}{\beta\hbar}  
\end{equation}
\vspace{5mm}
\subsection{Proof of Eq. \ref{eq:translation}}
\label{sec:translation} 
The momentum distribution of a system of $N$ particles in $d$-dimension  is 
\begin{align*}
  n(\vec p) & = \braket{\delta(\hat{\vec p} - \vec p)} = \braket{\frac{1}{(2\pi)^{Nd}}\int d\vec x e^{i(\hat{\vec p} - \vec p) \cdot \vec x}}
\\
&= \frac{1}{(2\pi)^{Nd}}\int d\vec x e^{-i\vec p \cdot \vec x}\braket{e^{i\hat{\vec p} \cdot \vec x}}
\\
&= \frac{1}{(2\pi)^{Nd}}\int d\vec x e^{-i\vec p \cdot \vec x}\braket{\hat{T}_{\vec x}}
\end{align*}
where $\hat{T}_{\vec x}$ is the translation operator by displacement $\vec x$.   
We thus identify the end-to-end distance distribution with the quantum-statistical average of the translation operator:  
\begin{equation}
  \tilde{n}(\vec x) = \frac{\Tr{(\hat{T}_{\vec x} \hat{\rho})}}{Z} = \sum_{n=1} \rho_n \int d\vec r d\vec r' \psi_n^*(\vec r') \psi_n(\vec r + \vec x)
\end{equation}
\bibliographystyle{apsrev}
\bibliography{abc}
\end{document}